\documentclass[prl,superscriptaddress,twocolumn,footinbib]{revtex4}
\pdfoutput=1
\usepackage[colorlinks=true,urlcolor=blue]{hyperref}
\usepackage{amsfonts}
\usepackage{graphicx}
\usepackage{placeins}
\usepackage{amsmath}
\usepackage{xcolor}
\usepackage{bm}
\usepackage{url}

\usepackage{color}
\definecolor{red}{rgb}{0.75,0,0}
\definecolor{blue}{rgb}{0,0,0.75}
\definecolor{green}{rgb}{0,0.5,0}

\DeclareMathOperator{\erf}{erf}
\DeclareMathOperator{\erfc}{erfc}

\begin{document}
	
\title{How to puncture a biomembrane: elastic versus entropic rupture}
\author{R. Capozza}
\affiliation{Institute for Infrastructure and Environment, School of Engineering, The University of Edinburgh, Edinburgh EH9 3JL}
\affiliation{Istituto Italiano di Tecnologia (IIT), Via Morego 30, 16163 Genova, Italy}
\author{L. Giomi}
\affiliation{Instituut-Lorentz, Universiteit Leiden, P.O. Box 9506, 2300 RA Leiden, The Netherlands}
\author{C. A. Gonano}
\affiliation{Politecnico di Milano, Via G. La Masa, 34, 20156 Milano, Italy} 
\author{F. De Angelis}
\affiliation{Istituto Italiano di Tecnologia (IIT), Via Morego 30, 16163 Genova, Italy}

\begin{abstract}
A very common strategy to penetrate the cell membrane  and access the 
internal compartment, consists of using sharp tips or nano needles.
However recent experiments of cell penetration by atomic force
microscopy tips show, contrary to expectations, a weak dependence of
penetration force on the curvature of the tip. Using molecular 
dynamics simulations and analytical arguments, here we show that 
membrane disruption can be driven either by elastic or entropic 
forces depending on the membrane size. Our findings have potentially 
relevant implications in tissue engineering and drug delivery, as 
they help assessing the effectiveness of the most common membranes 
penetration methods.
\end{abstract}

\maketitle

Lipid membranes are among the most versatile components of the cell \cite{alberts02,gennis89}. 
On the one hand, they act as scaffolds for countless biochemical reactions involving 
membrane-associated proteins, ribosomes etc. On the other hand, they protect the internal 
organelles from mechanical and chemical stimuli, by confining the cytoplasm and keeping 
in-and-out trafficking under strict regulation. Gaining access to the interior of the 
cell by penetrating the membrane is, therefore, of fundamental importance for many biological 
processes and applications such as electrical recording \cite{dipalo18,tian19}, drug and 
bio-molecular delivery \cite{ma12,xie12,chen19}.
%
%

Common strategies to investigate cell membrane penetration rely on atomic force microscopy (AFM) and nanowires arrays, complemented by theoretical models based on continuous elasticity \cite{xie13,xie15}. In the case of cell penetration by sharp tips, these models predict a strong dependence of the penetration force on the radius of curvature of the tip: the lower the radius of curvature the higher the local pressure, making penetration likely even at low forces. This is consistent with everyday-life experience: to punch a balloon one would rather use a sharp pin than a blunt one. Yet, recent results from AFM experiments show a radically different and unexpected behavior. The force exerted to penetrate the cell depends only weakly on the radius of curvature of the tip over more than one order of magnitude \cite{angle14,obataya05}. In particular, Angle {\em et al}. \cite{angle14} have demonstrated that penetration forces measured with tips having radius of curvature $20$ nm and $300$ nm differ only by 40\%, despite the corresponding pressure changes by more than a factor twohundred! Similarly, Obataya {\em et al.} \cite{obataya05} reported larger penetration force for pyramidal tips than for cylindrical ones, regardless the obvious difference in sharpness.

In this Letter we aim at clarifying this apparent contradiction. 
Using molecular dynamics simulations (MD) and Helfrich's continuous 
theory, we demonstrate that the rupture of a lipid membrane indented 
by a microscopic tip originates from two types of forces: short-ranged 
contact forces, exerted by the tip upon the elastically deformed 
membrane, and long-ranged entropic forces, resulting from the 
confinement of the thermal undulations. Depending on the size of the 
membrane, hence the amplitude of thermal undulations, entropic 
rupture can occur before indentation, thus making the sharpness of 
the tip unimportant.   

%

\begin{figure}[t]
\centering
\includegraphics[width=1.\linewidth]{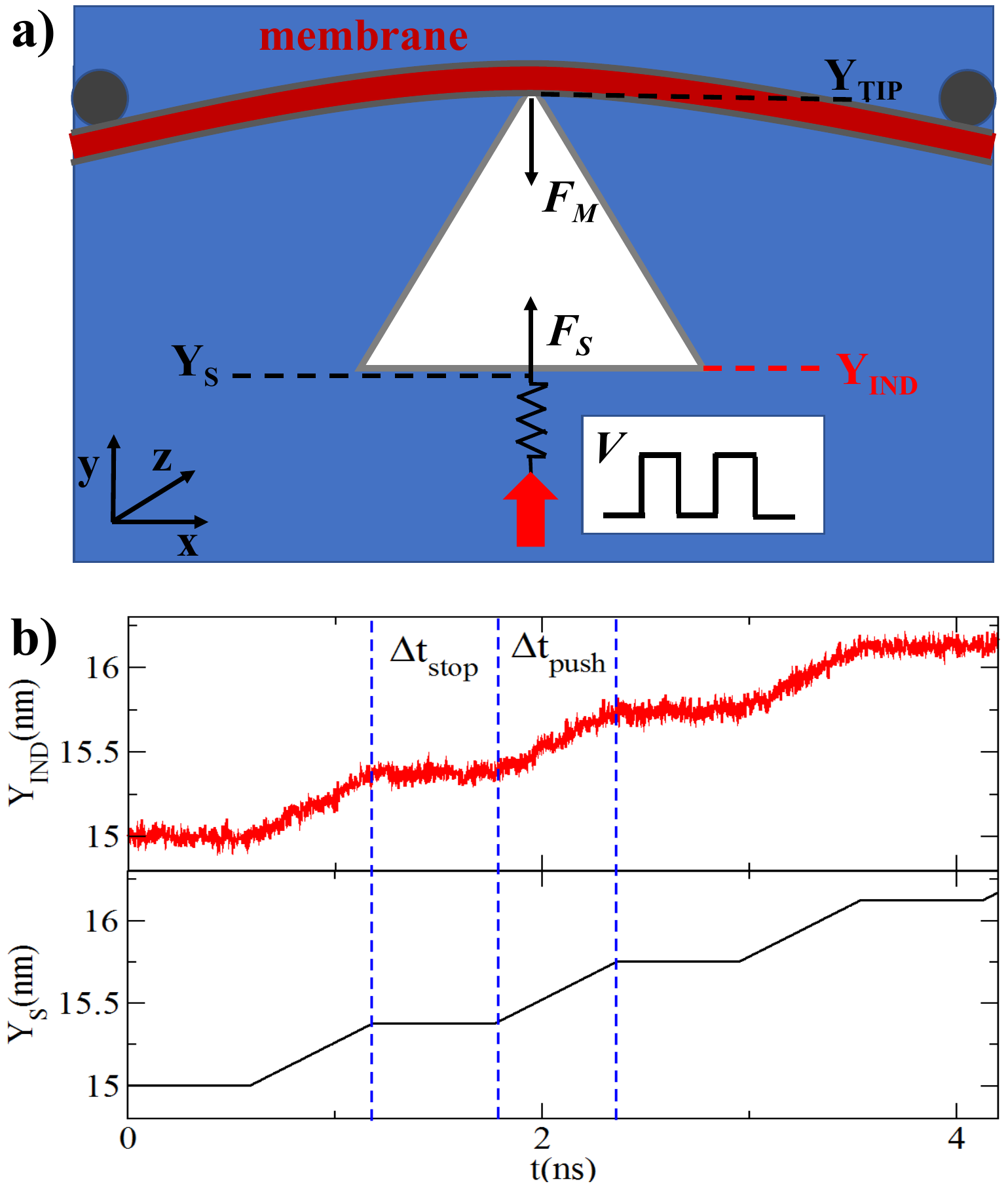}
\caption{(a) Schematic representation of the computational setup used in this study. (b) Stepwise simulation protocol to measure the break-through force. $Y_S$ and $Y_{\rm IND}$ are the vertical coordinates of spring and indenter respectively. The spring is pushed at velocity $V=0.64$~m/s for a time $\Delta t_{\rm push}=0.6$~ns, then it stops for a time $\Delta t_{\rm stop}=0.6$~ns.}
\label{fig1}
\end{figure}

Our MD simulations are based on a two-dimensional coarse-grained model, 
in which lipids are represented as a single hydrophilic Lennard-Jones 
(L-J) bead (i.e. the head) connected to a chain of five hydrophobic L-J 
beads (i.e. the tail). The beads have diameter $d_{0}=0.5$ nm. 
Water molecules are also modeled as L-J beads \cite{venturoli06,ayton06}. When lipid molecules are randomly dispersed in water they self-assemble 
in the form of a bilayer, with the hydrophilic heads pointing toward 
water and the tails clustering together to shield from the solvent 
\cite{SI}. The membrane is tensionless and its average conformation is 
initially straight. The area expansion modulus, $k_{a}$, and
the bending stiffness, $k_b$, 
of this model bilayer have been already estimated 
elsewhere \cite{capozza18}. The indenter consists of a rigid cluster 
of repulsive L-J particles connected to a spring of stiffness $K$. From 
the compression/elongation of the spring we obtain a direct measure 
of the force on the indenter (Fig.\ref{fig1}a). Evidently, our 
two-dimensional model cannot reproduce all the processes that occur 
in three-dimensional systems, such as lipid diffusion through the 
membrane, but represents a good compromise between the accuracy and 
computational cost and allows a statistical analysis of the results.
The maximum size $L$ of our system is $235$ nm and periodic boundary conditions are applied. 

As the indenter approaches the lipids, it experiences a force $F_{M}$, due to the interaction with the membrane, and a drag force $F_D$ from the surrounding fluid. 
In order to quantify the latter, we have calculated the time-averaged force $\langle F_D \rangle$ on the indenter as a function of its velocity, $V$, without the membrane (for detail see Ref. \cite{SI}). In the range of parameters explored, the behavior of $\langle F_D \rangle$ versus $V$ is linear to a good approximation. This allows us to estimate the viscosity as $\eta_d = 8 \times 10^{-3}$~Pa~s.
More importantly, this calculation demonstrates that, even at the lowest velocity value accessible to our simulations, the drag force is $F_D \approx 7.2$~pN. This is comparable with the membrane tensile strength \cite{capozza18}, thus implies that the drag force exerted by the solvent is never negligible.
In order to circumvent these limitations and decouple $F_{M}$ from $F_{D}$, we opted for a stepwise loading protocol. We first pull the spring at constant speed, $V=0.64$~m/s, for a time $\Delta t_{\rm push}=0.6$~ns and then we pause for a time interval $\Delta t_{\rm stop}=0.6$~ns (Fig.\ref{fig1}b). During $\Delta t_{\rm stop}$ we measure the time-averaged force experienced by the spring $\langle F_{S}\rangle$. Fig.\ref{fig1}b shows a comparison between the position of the indenter, $Y_{\rm IND}$, and the spring, $Y_{S}$, versus time. The data demonstrate that our protocol is quasi-static and that, during the time interval $\Delta t_{\rm stop}$, the indenter fluctuates around its equilibrium position, thus validating the approximation $F_{M} \approx \langle F_{S} \rangle$.

%

\begin{figure}[t]
\centering
\includegraphics[width=1.\linewidth]{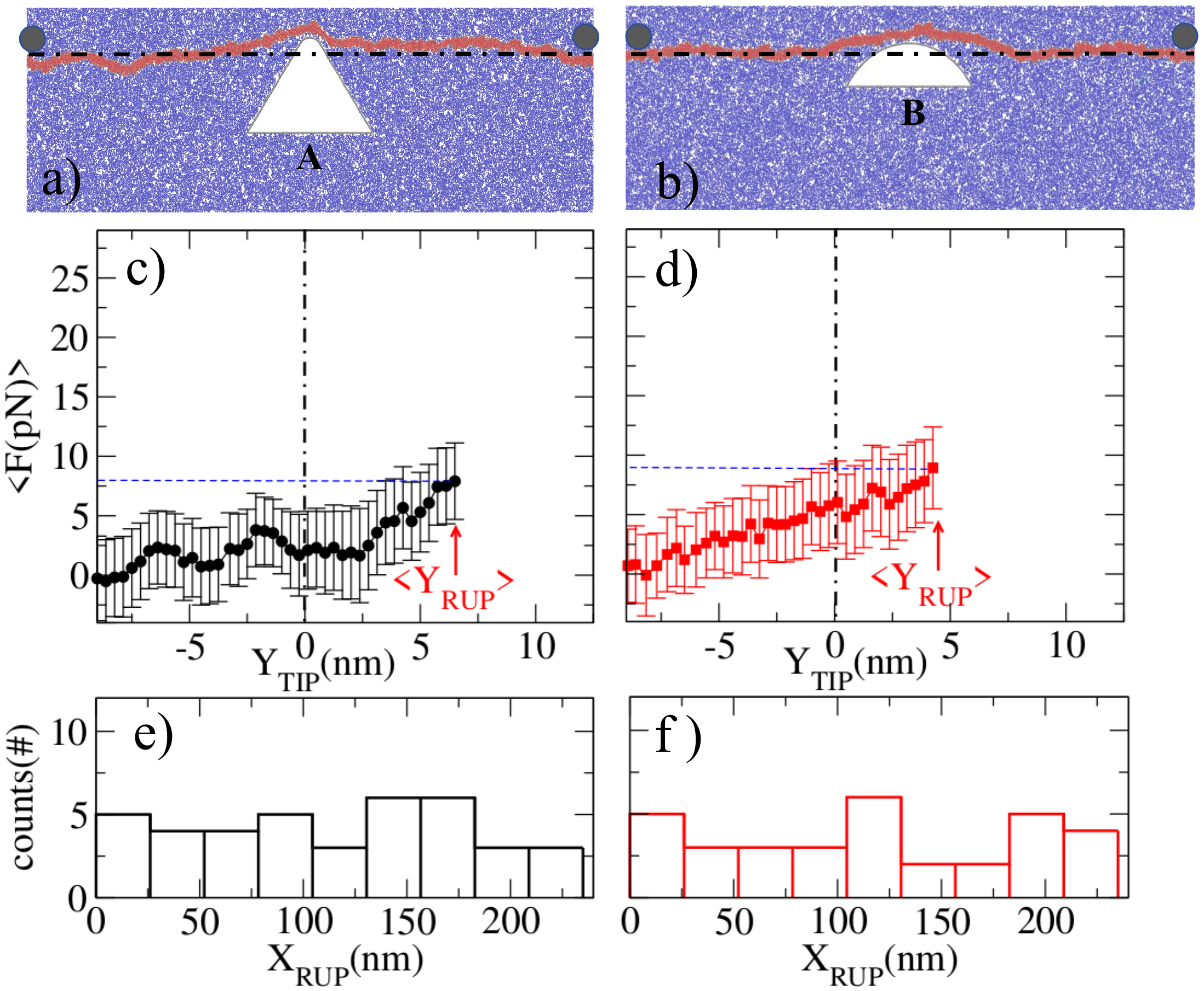}
\caption{Snapshots from simulations and loading setup. 
	The membrane is 
	loaded by two indenter, $A$ and $B$, with different radii of 
	curvature. Two cylindrical posts block the membrane at the border
	of simulation box (gray circles). The horizontal dotted dashed 
	line indicates the taut straight membrane. Force $F$ vs tip 
	position $Y_{\rm TIP}$ for the sharper (c) and flatter (d) tip. 
	Distribution of rupture positions $X_{\rm RUP}$ for the $A-$tip 
	(e) and $B-$tip (f). The membrane breaks at a random location, 
	uniformly distributed between the blocking posts. The membrane 
	size is $L=235$~nm.}
\label{fig2}
\end{figure}

In order to investigate how the sharpness of the indenter affects the penetration force, we have considered two indeters, labeled $A$ and $B$, having radius of curvature $R_{A}=5$~nm and $R_{B}=30$~nm respectively. The membrane is kept in place by a pair of cylindrical posts held at a distance $L$ from each other (see Fig.\ref{fig2}a,b). In this configuration, resembling the set-up of a pore-spanning membrane loaded by an AFM tip \cite{steltenkamp06,mey09}, the system is free to vibrate as a result of thermal motion of the lipids. Fig.\ref{fig2}c,d shows the force experienced by the indenter as a function of its vertical position $Y_{\rm TIP}$. The data points and error bars are obtained by averaging over 20 different simulations. Remarkably, even when the tip of the indenter is far from the membrane (i.e. $Y_{\rm IND} \approx -7$~nm, well below the dashed line in Fig.\ref{fig2}a,b indicating the membrane straight profile), we detect a measurable force as reported experimentally \cite{alessandrini12}. Since for $V=0$ there are no other forces acting on the membrane except the contact force exerted by the indenter and those arising from thermal fluctuations, we ascribe the origin to the signal measured in the absence of contact to entropic forces. 

In addition to the force, we measure the position point of rupture: $(X_{\rm RUP},Y_{\rm RUP})$. 
In particular, the average vertical position $\langle Y_{\rm RUP} \rangle$ of the rupture point, 
corresponding to maximal indentation depth, is marked by a red arrow in Fig.\ref{fig2}c,d.
These results clearly show that both the maximal indentation depth and the rupture force $\langle F(Y_{\rm RUP})\rangle$ are comparable for the two indenters, despite the difference in sharpness and consistent with AFM experiments \cite{angle14,obataya05,obataya05}.
As a further demonstration of the irrelevance of the tip radius of curvature, we report in Fig.~\ref{fig2}e,f the distribution of the horizontal position $X_{\rm RUP}$ of the point of rupture for the $A-$ and $B-$tip. Even for the sharper $A-$tip, the membrane breaks at a random location, uniformly distributed between the two blocking posts and not at the center, as one would expect by viewing the membrane as an elastic sheet punctured in the middle by a sharp pin. 

%
%
\begin{figure}[t]
\centering
\includegraphics[width=1.\linewidth]{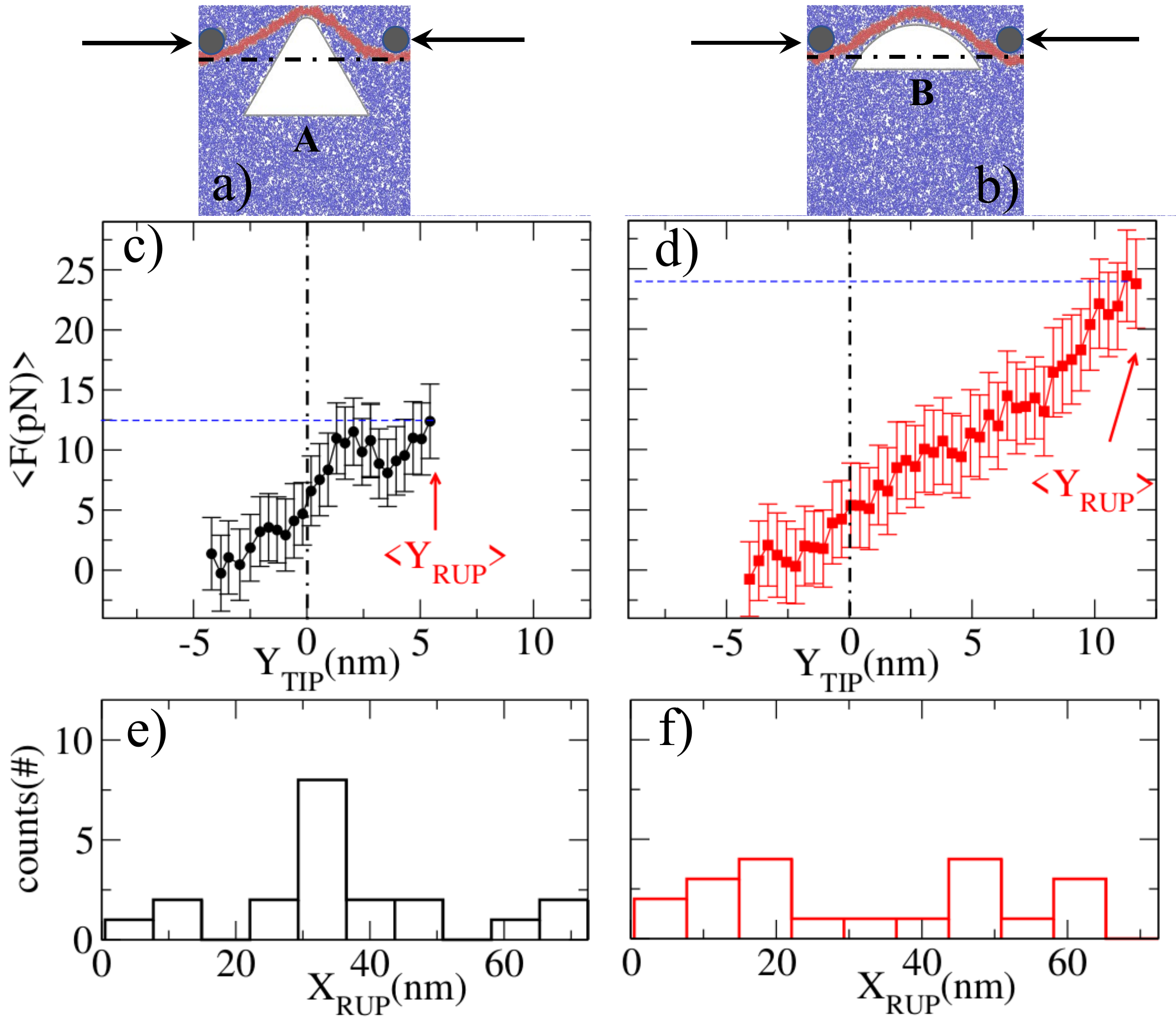}
\caption{(a,b) Snapshots from the MD simulations. Force $F$ vs tip 
	position $Y_{\rm TIP}$ for a sharp (c) and smooth (d) tip and
	a membrane size $L=72$~nm. 
	Distribution of rupture positions $x_{\rm RUP}$ for the $A-$tip 
	(e) and $B-$tip (f). When the $A-$tip is used, the membrane 
	shows a preference to break in the center, at the apex.}
\label{fig3}
\end{figure}
The situation drastically changes when the system size is reduced to $L=72$~nm, as shown in Fig.\ref{fig3}. 
The maximal indentation depth $\langle Y_{\rm RUP}\rangle$ and the rupture force $\langle F(Y_{\rm RUP}) \rangle$ 
are prominently smaller for the sharper $A-$tip than for the flatter $B-$tip, as expected from elastic materials. 
Furthermore the distribution of $X_{\rm RUP}$, shown in Fig.\ref{fig3}e,f, indicates now a preference towards 
the center in the case of the $A-$tip. These results demonstrate that, 
upon reducing the thermal undulations of the membrane by shortening the system size $L$, 
the system crossovers from an entropic to an elastic regime, in which rupture results as 
a direct consequence of the applied pressure. A similar trend is also obtained by keeping 
the system size to $L=236$~nm but moving closer the blocking posts at distance $D=72$~nm (see Ref. \cite{SI} 
for more details). 
   
Entropic forces in confined lipid membranes, also known as undulation forces, have been predicted four decades ago by Helfrich in the context 
of multilayered structures \cite{helfrich78} and later investigated 
experimentally and theoretically \cite{abillon90,israelachvili11,lindahl00,ayton06}.
Analogously to other types of entropic forces, undulation forces arises when spatial confinement limits the fluctuations of the membranes, thus leading to a decrease in entropy. In order to illustrate this concept, in the following we calculate the force experienced by an effectively one-dimensional membrane loaded by an infinitesimally sharp tip. Under the assumption of small fluctuations, the membrane's free-energy can be expressed as (see e.g. Ref. \cite{israelachvili11}):
\begin{equation}\label{eq:helfrich}
F = \frac{1}{2}\int_{-\frac{L}{2}}^{\frac{L}{2}} {\rm d}x\,\left[a(h')^{2}+b(h'')^{2}\right]\;,		
\end{equation}
where $h=h(x)$ is the height of the membrane above the $x-$axis of standard 
Cartesian frame, $a= k_{a}d_{0}$ and $b = k_{b}d_{0}$ are, respectively, the effective bending stiffness 
and surface tension and the prime indicates differentiation with respect to $x$.

\begin{figure}[t]
  \centering
  \includegraphics[width=1.\linewidth]{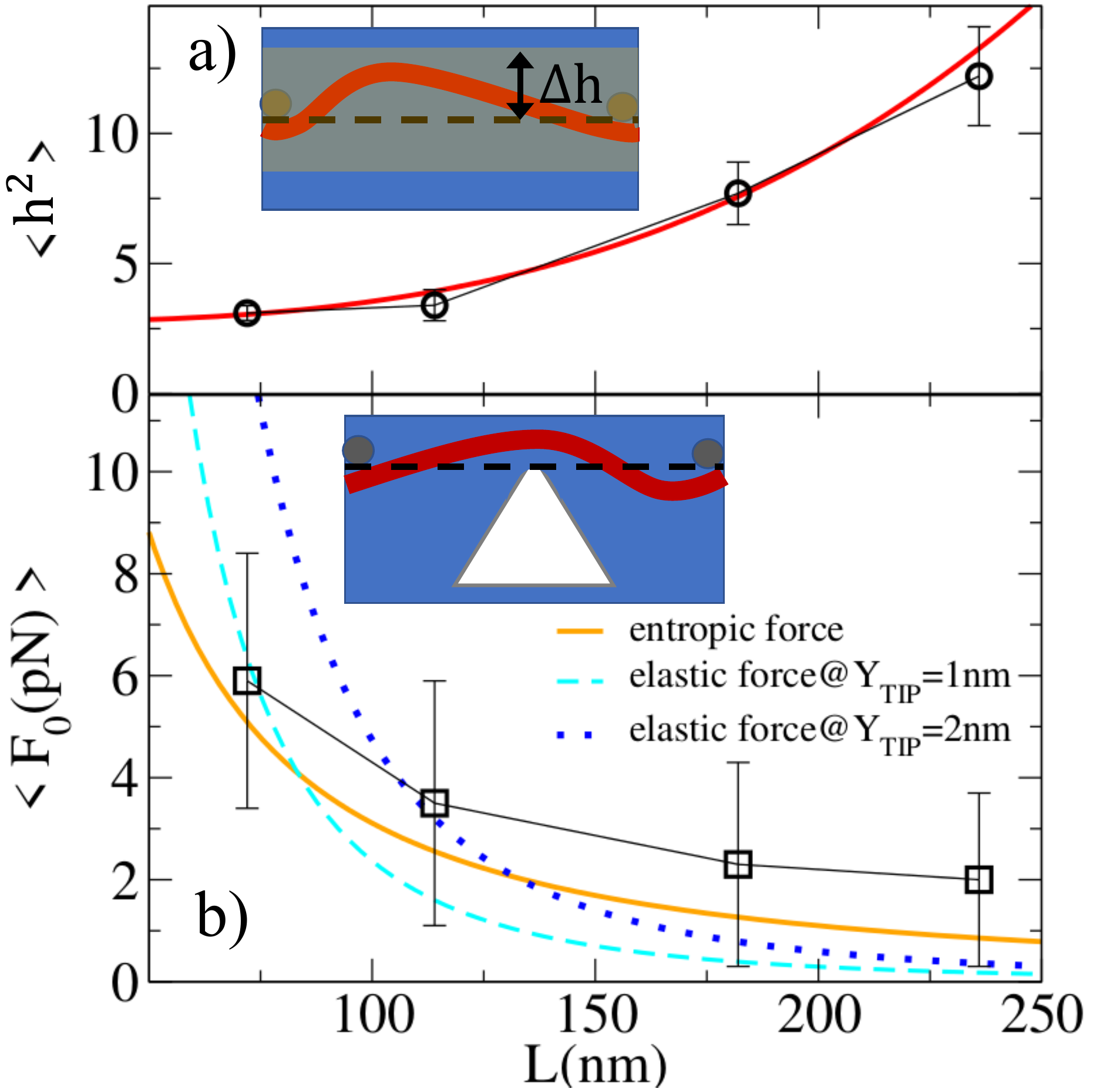}
	\caption{a) Average squared amplitude $\langle h^2 \rangle$ 
        of fluctuations vs membrane size $L$. The black circles
	are the results of simulations, compared with the theoretical 
	red line from Eq.~\eqref{eq:ampl}.
	The shaded yellow area in the inset indicates 
	the average amplitude of membrane fluctuations $\Delta h = \sqrt{\langle h ^{2} \rangle}$. 
	Thermal fluctuations are suppressed as the membrane is shortened. 
	b) Force $F_0$ vs $L$ experienced by the sharp indenter while 
	kept at fixed position $Y_{\rm TIP}\approx 0$, when the elastic term 
	becomes negligible. 
	The continuous orange line is the the entropic term in 
	Eq.\eqref{eq:fa}, black squares the 
	corresponding estimates from simulations. The dashed and dotted 
	lines are the values of elastic forces at $Y_{\rm TIP}=1$~nm and 
	$Y_{\rm TIP}=2$~nm. As $L$ is reduced, the elastic forces 
	at 1 nm and 2 nm overtake 
	the entropic component and become more and more separated, 
	indicating an increase of the rate with which the force grows.}
  \label{fig4}
\end{figure}

First, we show that the thermal undulations of the membrane increases 
with the system size, thus rendering large membranes more prone to 
rupture than smaller membranes. This is readily done upon expanding 
$h$ in Fourier modes, i.e. $h(x)=(1/L)\sum_{q}h_{q}\,e^{iqx}$, 
with $q=\pm 2\pi/L,\,\pm 4\pi/L\ldots\,\pm 2\pi/d_{0}$. 
Now, in our system 
$L\gg d_{0}$ and $a R^{2}/b \ll 1$, with $R$ the typical radius 
of curvature of the membrane. Then, approximating Eq. \eqref{eq:helfrich}
as $F\approx b/L\sum_{q=2\pi/L}^{\infty}q^{4} h_{q}h_{-q}$ and using 
the equipartion theorem yields: $\langle h_{q}h_{-q}\rangle=k_{B}TL/(bq^{4})$, from which 
\begin{equation}\label{eq:ampl}
\langle h^{2} \rangle = \frac{2}{L^{2}}\sum_{q=2\pi/L}^{\infty}\langle h_{q}h_{-q}\rangle=\frac{k_{B}TL^{3}}{720\,b}\;.
\end{equation}
Thus, the membrane mean squared height scales as $\langle h^{2} \rangle \sim L^{3}$ with the length $L$ of the membrane, in good agreement with the result of our MD simulations, 
as shown Fig.\ref{fig4}a. Analogously, one can estimate the radius 
of curvature of an undulation mode of amplitude 
$\Delta h=\sqrt{\langle h^{2} \rangle}$, namely:\cite{SI} 
\begin{equation}\label{eq:radius}
R \sim [b^{2}\Delta h / (k_{B}T)^{2}]^{1/3} 
\end{equation}
Next, following Daniels and Turner \cite{daniels2004}, we calculate the 
partition function of a membrane whose mid-point is constrained to 
fluctuate only above a certain height, i.e. $h(0) \ge Y_{\rm TIP}$, 
representing the position of the tip of an infinitesimally sharp indenter. 
This is given by:
\begin{equation}\label{eq:z}
Z_{\rm TIP} = \int_{Y_{\rm TIP}}^{\infty} {\rm d}y\,\int Dh\,\delta[h(0)-y]\,e^{-\frac{F}{k_{B}T}}\,.	
\end{equation}
The path integral over the configuration of the height field $h$ is readily calculated using 
standard algebraic manipulations. This yield, up to a constant prefactors, $Z_{\rm TIP} = \erfc[Y_{\rm TIP}/(\sqrt{2}\,\Delta h)]$, where $\erfc(\cdots)=1-\erf(\cdots)$ is the complementary error function. Differentiating the free energy $F_{\rm TIP}=-k_{B}T\log Z_{\rm TIP}$ with respect to $Y_{\rm TIP}$ and expanding for around $Y_{\rm TIP}=0$ finally yields an expression for the force experienced by the membrane in proximity of the tip, namely: 
\begin{equation}\label{eq:fa} 
f \approx \frac{1440\,b}{\pi L^{3}}\,Y_{\rm TIP}+\sqrt{\frac{1440\,b\,k_{B}T}{\pi L^{3}}}\;,  
\end{equation}
A full derivation of Eq. \eqref{eq:fa} is reported in Ref. \cite{SI}. 

Some comments are in order. Eq. \eqref{eq:fa} consists of two terms representing the elastic and entropic contributions respectively. At $T=0$, the second term vanishes and the force matches the Euler-Bernoulli loading force in a three-point bending configuration (see \cite{SI} and Ref. \cite{timoshenko12}). 
More importantly, Eq.~\eqref{eq:fa} demonstrates that upon increasing 
$L$, the force crosses over from elastic (small $L$) to entropic 
(large $L$). 

To highlight the role of the entropic contribution, we have used MD simulations to calculate the force experienced by the $A-$indenter at $Y_{\rm TIP}\approx 0$, when the elastic contribution becomes negligible. 
This is shown in Fig.~\ref{fig4}b as a function of the membrane size $L$ (square dots), together with the analytical estimate given in Eq. \eqref{eq:fa} (solid line).  
By constrast, the dashed and dotted lines show the elastic contribution to the force given in Eq. \eqref{eq:fa} at the same temperature, but for $Y_{\rm TIP}=1$~nm and $Y_{\rm TIP}=2$~nm respectively. As $L$ is reduced, the elastic contribution overweights the entropic one even at small deformations. 

Combining these findings with the results of our MD simulations, 
we conclude that the rupture of a lipid membrane indented by a 
microscopic tip results from two different forces: a standard Hookean 
force, originating from a direct contact between the tip and the 
membrane and linearly proportional to the indentation depth, and an 
entropic force, determined the hindrance of the thermal ondulation
caused by the presence of the tip. Depending on the size of the 
membrane, hence the magnitude of the height fluctuation $\Delta h$, 
and the indentation depth $Y_{\rm TIP}$, rupture can occur as a 
consequence of the elastic force, for $Y_{\rm TIP} \gg \Delta h$, or 
as a consequence of the entropic force, for $Y_{\rm TIP} \ll \Delta h$.
In this latter case, the radius of curvature 
of the membrane does not depend on the sharpness of the 
indenting tip, but is determined by the temperature, $T$,
bending stiffness, $k_b$ and mean amplitude of fluctuation, $\Delta h$, 
as shown in Eq.\eqref{eq:radius}.

Furthermore, we stress that, at constant indentation velocity, 
large membranes are more prone to entropic rupture and lower 
rupture forces. In fact, as shown in Eq. \eqref{eq:fa}, the loading rate 
(i.e. the rate with which the force grows) decreases as the membrane 
size increases.
Considering that the rupture force of biomembranes grows 
logarithmically with the loading rate \cite{evans03}, we expect
a decrease of the rupture force in the entropic regime where 
the loading rate is low. This is in agreement with the
simulation results of our computer model, showing indeed 
loading rate-dependent rupture forces \cite{SI}.

Our findings have a potential immediate application to tissue 
engineering and drug delivery, where, in order to achieve a fast 
and effective access to the cell's interior, nanometer-sized needles 
are often employed \cite{shalek10,gopal19}. Our analysis suggests that these 
techniques can be optimized by suppressing the thermal undulations, 
as in the case of cells cultured on sharp nanopillars. 
In this set-up, the cell tightly 
wraps around the pillar \cite{berthing12,zhao17}, which then determine 
the shape of the plasma 
membrane, thus hindering thermal fluctuations. Conversely, when cells 
are cultured on a substrate, the free surface in contact with the 
culture medium is mobile and subject to thermal fluctuations \cite{evans08}. Such a 
mobile, undulating interface could be insensitive to the geometry and 
sharpness of an approaching AFM tip, thus rendering standard perforation 
strategies uneffective.





%
%
\bibliography{biblio}

\end{document}